\documentstyle[psfig]{mn}
\input{psfig.sty}

\newif\ifAMStwofonts

\title[Winds from massive stars and the afterglows of
      $\gamma$-ray bursts]  
      {Winds from massive stars: implications for the afterglows of
      $\gamma$-ray bursts}  
\author[E. Ramirez-Ruiz et al.]
       {Enrico Ramirez-Ruiz$^{1}$, Lynnette M. Dray$^{1}$, Piero Madau$^{1,2}$
and Christopher A. Tout$^{1}$
\\${\bf 1.}$ Institute of Astronomy, Madingley Road, Cambridge, CB3 0HA.
\\${\bf 2.}$ Department of Astronomy and Astrophysics, University of
      California, Santa Cruz, CA 95064, USA.}

\date{}

\begin{document}

\maketitle

\label{firstpage}

\begin{abstract}

Recent observations suggest that long-duration $\gamma$-ray bursts
(GRBs) and their afterglows are produced by highly relativistic jets
emitted in core-collapse explosions. The pre-explosive ambient
medium provides a natural test for the most likely progenitors of
GRBs. Those stars that shed their envelopes most readily
have short jet crossing times and are more likely to
produce a GRB.  We construct a simple computational scheme to
explore the expected contribution of the  presupernova
ejecta of single Wolf-Rayet (WR) stars to the
circumstellar environment. Using
detailed stellar tracks for the evolution of massive stars, we discuss
the effects that the initial main sequence mass,
metallicity, rotation and membership in a binary system have
on the ambient medium. We extend
the theory of GRB afterglows in winds to consider the effect of the
relativistic fireball propagating through the WR ejecta. Specific
predictions are made for the interaction of 
the relativistic blast wave with the density bumps that arise when
the progenitor star rapidly loses a large 
fraction of its initial mass or 
when the ejected wind interacts with the
external medium and decelerates. A re-brightening of the afterglow with
a spectrum redder than the typical synchrotron spectrum (as seen in
GRB 970508, GRB 980326 and GRB 000911) is predicted. We also calculate the 
luminosity of the reflected echo that arises when circumstellar
material Compton-scatters the prompt radiation, and examine
the spectral signatures expected from the interaction of the GRB
afterglow with the ejected medium. 
 
\end{abstract}


\section{Introduction}

Recent evidence has given support to the idea that long-duration  $\gamma$-ray
bursts (GRBs) result from the cataclysmic
collapse of massive stars in very energetic supernova-like explosions
\cite{mc01} rather
than the coalescence of two compact objects,
such as black holes or neutron stars \cite{lat76}. The observed 
distribution of optical afterglow luminosities with respect to their
host galaxies   
suggests that some GRBs may be associated with star-forming regions
\cite{pac98}. The spatial and temporal coincidence between the supernova
(SN) 1998bw and GRB 980425 \cite{gal98} hints that some GRBs may be linked to
ultra-bright type-Ibc SNe but this is
still controversial \cite{pian00}. Some evidence for
supernova-type emission has also been found in GRB 980326
\cite{blo99}, GRB 970228 \cite{dan99} and GRB 000911 (Lazzati et
al. 2001). If we  believe that GRBs  are associated
with SN-like events, then the explosion is likely to be  jet-induced 
if the baryon contamination problem is to be avoided
\cite{Rees92}. Without collimation, the maximum isotropic 
$\gamma$-ray emission is raised to unprecedented values, implying a
total energy of the order of $3 \times 10^{54}$ ergs (GRB 990123;
Kulkarni et al. 1999). A beamed jet alleviates this energy
requirement. The evidence is still preliminary, but collimation
factors of $\Omega_j /4\pi <0.01$ have been derived from the decline in some
afterglow light-curves \cite{kul99,castro99,har99}. Whether a jet is
present or not, the energy budget is compatible with
current scenarios that involve the collapse of compact objects:
the binding energy of 
the orbiting debris and the spin energy of the black hole are the two
main energy reservoirs available \cite{Rees99}.\\ 

MacFadyen \& Woosley (1999; hereafter MW99)
have explored the evolution of rotating helium stars whose iron core
collapse does not produce a successful traditional neutrino-powered
explosion. For massive stars ($M_{0} \ga 35 {\rm M}_{\odot}$)
with sufficient angular momentum, a rapidly accreting stellar mass
black hole forms promptly at the centre of the collapsing star (Type I
collapsars). Less rapidly accreting black holes can also
form over longer timescales due to the infall of stellar material
that failed to escape during the initial SN (Type II
collapsars). Prompt and delayed black hole formation can occur
in stars with a range of radii depending on the evolutionary
state of the massive progenitor, its metallicity and its membership in a binary
system. If the star loses its hydrogen envelope along the way, and if
the jet produced by the accretion maintains its energy for longer
than it takes the jet to tunnel through the star, a
GRB is likely to be produced (MW99). Otherwise, acceleration
of the explosion debris to a sufficiently high Lorentz factor ($>10^{2}$,
M\'esz\'aros \& Rees  1997) is unlikely and an asymmetric supernovae
like SN 1998bw may result \cite{mc01}. \\

In the collapsar model for GRBs,
or in any other model involving a massive star, the key to
obtaining relativistic motion is the escape 
of an energy-loaded fireball from the stellar environment. This is
aided if the
progenitor undergoes a Wolf-Rayet (WR) phase, which is
characterized by a strong stellar wind that causes the
star to lose enough of its outer layers for the surface  hydrogen
abundance to become minimal. The radius of a WR star is
sufficiently small for the explosion energy to break out
before the engine ceases to operate. 
Because of the intrinsic variations of mass-loss rates in WR evolution, the 
GRB blast wave expands into shells of varying gas density.
The effects of the WR ejecta interacting with the interstellar medium (ISM)
can be observed in the wind bubbles around
some of these objects. The
deceleration of a pre-SN wind by the pressure of the surrounding 
medium  or the interaction of fast
and slow winds (Luo \& McCray 1991; Vikram \& Balick 1998) 
could also create circumstellar shells.\\

It is an unavoidable consequence of a core collapse model
for GRBs  that the afterglow emission propagates  
in a dense stellar wind. Chevalier \& Li (1999, 2000)
described some features of the afterglow evolution in a steady,
spherically symmetric wind with density  $\propto r^{-2}$. In this
paper we study the effects of strong, non-steady WR 
stellar winds with large mass-loss variations and non-spherical
geometries. In particular  we explore the 
impact of the pre-SN
evolution of single high-mass stars\footnote{The comparative
modelling of single and binary systems will be presented in a
subsequent paper (Dray et al. 2001 in preparation).}
($M_{{0}} \ga 35 { \rm M}_\odot$) on the
circumstellar environment. In $\S$ 2 we use detailed stellar
evolution of WR stars to show how the progenitor initial mass ($M_{0}$) and
metallicity $Z$ affect the density profile of the ambient material. 
In $\S$ 3 we extend the theory of
afterglows in winds to consider the propagation of a relativistic
fireball through the dense
environment expected at the end of the evolution of a WR
star. We calculate the echo that arises
when the progenitor circumstellar material Compton-scatters the prompt
$\gamma$-ray radiation in $\S$ 4. Implications of recent observations,
in particular for GRB 970508, GRB 980326 and GRB 000911,
and predictions for future ones are discussed in $\S$ 5. We
examine the spectral line signatures expected from the interaction of
a GRB afterglow with the chemically enriched environment in $\S$ 6. Our
conclusions are presented in $\S$ 7.

\section{Evolutionary Models of Massive Stars with High Mass-Loss Rates }

\subsection{Model outline}
\subsubsection{WR evolution}
WR stars have many extreme properties: they have
luminosities typically between 10$^{5}$ and
10$^{6}$ ${\rm L}_\odot$ and are very hot. This places them in the
upper left corner of the HR diagram. Because of their dense stellar winds
($10^{-4} {\rm M}_\odot {\rm yr}^{-1}$), the hydrostatic surfaces of WR
stars are 
mostly inaccessible to direct observations. This  makes it
difficult to associate stellar models with observed objects on the
basis of their radii, surface temperatures or luminosities. The
spectroscopic determination of the abundances of H, He, C, N and O (e.g.
Smith \& Hummer 1998) allows only certain classes of
theoretical stellar models. Massive stars are
known to have stellar winds and mass-loss while on the main
sequence and, although the amount of mass lost here
may be small compared to that lost in later phases, it is 
relevant to the subsequent evolution. The first evolutionary phase of a
WR star is the nitrogen-line stage (WNL), which begins when CNO
processed material is exposed at the stellar surface. 
This progresses to the WNE 
stage during which no surface hydrogen is detectable. Once the
hydrogen envelope is completely lost, the WR models strictly obey the
mass-luminosity relationship for hydrogenless WR stars. This allows
the details of their further evolution to be calculated
by solving only the equation of global energy conservation
\cite{lan89} for a given mass-loss rate, because the mass determines
the luminosity which in 
turn determines the rate of nuclear burning in the stellar core. After
the helium envelope is shed, the
WC stage occurs, in which strong carbon lines can be seen. 
Finally, WO
stars are formed, with high surface oxygen abundances. 
Subclassifications in the above categories are made
according to line strength ratios (eg. Smith \& Maeder 1991).\\

In order to assess the simplest case, we shall confine the
discussion in this section to non-rotating, non-magnetic single
stars. We use detailed evolutionary tracks 
between the initial main sequence and the WO stage, with enhanced
mass-loss rates based on the work  
of  Meynet et al. (1994).
We carry out the computational simulations
with the  stellar evolution code first developed by Eggleton (1971) and 
recently updated by Pols et al. (1995). Due to
slight differences in  the stellar evolution codes, including a
different treatment of  convective overshooting \cite{schr97}, the
stellar tracks 
computed by  Meynet et al. (1994) are not identical to the ones
calculated here (e.g. the minimum initial mass for WR star formation is
slightly lower) but they are qualitatively the same.

As described in Meynet et al.
(1994), enhanced mass-loss rates were used during the phases when
stellar winds are believed to be driven by radiation
pressure, i.e. during the whole evolution with the exception of the
WNE and WC/WO phases, which correspond to the stage of
nearly pure helium envelopes and of He+C+O cores,
respectively. Enormous mass-loss can occur at the entrance into these
stages. Pre-WR mass-loss rates 
for the whole HR diagram were taken from the expression derived by de 
Jager  et al. (1988) and scaled with metallicity 
as $(\dot{ M_z}/\dot{\rm M}_{\odot}) =
(Z/{\rm Z}_\odot)^{0.5}$. For WNL stars a 
constant mass-loss rate was adopted \cite{abbo87} and for later  
stages of WR evolution a mass-dependent mass-loss rate was assumed as
specified by 
Langer (1989). In cases where the cumulative WNL
mass-loss was enough to nearly
evaporate the star we used the WNE rate instead.\\

\subsubsection{Wind-ISM interaction}

Winds from the most massive stars tend to have the greatest
effect on the ISM because the mass-loss rates are large and the
winds are very fast and
carry large momentum fluxes. Moreover, massive stars tend to lie in
associations,  so there is the cumulative influence of the winds of
many stars. In describing 
the effects of stellar winds on the surrounding medium we used the
basic theory for the wind-ISM interaction develop by Castor, McCray
$\&$ Weaver (1975). We assume
that the ISM has a simple structure, with no clumps, no magnetic
fields, no multiple ionization phases and no nearby 
stars. During 
the evolution of a wind-driven circumstellar shell the
system has a four-zone structure (analogous to that of a supernova
shell; Woltjer 1972; Cox 1972). From the inside to the outside these
zones are: (I) a supersonic stellar wind with density $\rho (r)=\dot{M}/ 4\pi
r^2 v_{\infty}$; (II) a hot, almost isobaric region consisting of shocked
stellar wind mixed with a small fraction of the swept-up interstellar
gas; (III) a thin, dense, cold shell containing most of the swept-up
interstellar gas; (IV) ambient interstellar gas of number density $n_{0}$.\\

The wind initially expands unopposed into the ISM with a 
velocity of about $v_{\infty}$, the escape velocity
at the sonic point (for the purpose of this analysis we assume
$v_{\infty}= \sqrt{2 G M_*/R_*}$, because the sonic point is
reached after a few stellar radii). The free expansion phase is considered to
be terminated at a time $t_{\rm dec}$, when the swept-up mass of the
interstellar medium is
comparable to the mass in the wind (Woltjer 1972; Castor et al. 1975;
McCray 1983). The mass lost by the star is 
$\dot{M}t_{ \rm dec}$ and the swept-up mass is $
{4\pi \over 3}(v_{\infty}\;t_{ \rm dec})^3 n_{0} m_{\rm p} \mu_c$, where
$\mu_c \sim$ 2 in a helium gas.  These two masses are equal when
$t_{\rm dec}=\sqrt{3\dot{M}/ (4 \pi v_{\infty}^3 n_{0} m_{\rm
p} \mu_c) }$, which is about 100 years for a typical WR  wind expanding into an
homogeneous ISM\footnote{The free expansion phase takes place at the early
stages of the evolution of the hot star and occupies a minimal fraction
of its lifetime. During this time both $\dot{M}$
and $v_{\infty}$ are approximately constant (see Figs. 1a and 1c).}. During
this time the wind bubble has reached a radius of 
\begin{equation}
r_{\rm dec}=8.8\times 10^{17} \dot{M}_{-5}^{1/2}
n_{0,0}^{-1/2}v_{\infty,3}^{-1/2} {\rm cm}, 
\end{equation}
where $\dot{M}$ is the mass loss rate in units of
solar masses per year, $n_{0}$ is the density of the surrounding
medium in units of cm$^{-3}$, $v_\infty$ the wind velocity in units of
km~s$^{-1}$ and we adopt the convention $Q = 10^x\,Q_x$.

In a dense ($n_{0} \approx 10^5 {\rm cm}^{-3}$) molecular cloud a
wind-driven bubble will radiate its energy and stall in a short time,
causing the ejected wind to slow rapidly at a much smaller 
radius (Shull 1982). The same may be true if a neighbouring star is
present - for example, a very close binary in which the WR wind is
strongly slowed down by the OB star radiation pressure before it
collides with the OB wind. $R_{\rm OB}<r_{\rm OB} \le r_{\rm ter}$,
where $r_{\rm OB}$ is the distance from the OB star (with radius
$R_{\rm OB}$) to the region 
of the stellar wind collision and $r_{\rm ter} \approx 5\;R_{\rm OB}$ is the
distance from the OB
star centre at which the OB wind reaches terminal velocity\footnote{
If a very close binary is present, the high density region will
probably be reduced to a small covering factor surrounding the
collapsing star. Note that a beamed fireball would introduce an
additional geometrical factor.}. Many WR
stars with absorption lines appear to be 
single; however, the fraction of visible close WR + OB binaries
seems to be around 35 per cent \cite{rev86}.\\

When the free expansion phase (I) has
ended, the wind encounters an inward facing shock. 
Kinetic energy is
deposited in the shocked wind region in the form of heat, 
\begin{equation}
T_{\rm shock}= {3 \over 16}{m_{\rm p} \mu_c \over k}(\Delta v)^2= 1.4 \times
10^5 ({\Delta v \over 100\;{\rm km\;s}^{-1}})^2 {\rm K},
\end{equation}
where $\Delta v$ 
is the relative speed of the material approaching the shock. This creates a
temperature of about $10^{7} {\rm K}$ in the shocked wind region. During
phase II, the material is so hot that it causes the contact surface
to move outward more slowly than it would in a freely expanding wind. 
The ISM that enters the outward facing shock is heated to a temperature below
$10^6 {\rm K}$, emission of line radiation becomes the dominant
cooling process and the swept-up gas cools quickly to temperatures of
about $10^4 {\rm 
K}$ that can be maintained by the radiation field of the star. The
duration of the adiabatic expansion phase can thus be estimated by
finding the time it takes the expanding gas to cool from $T_{\rm shock}
\approx 10^{7} {\rm K}$ to  $10^{6} {\rm K}$. Using equation (2),
we find that a change in temperature from  $10^{7} {\rm
K}$ to  $10^{6} {\rm K}$ corresponds to a change in jump velocity by a
factor of $\sqrt{10}$.  This change in jump velocity represents 
a  phase II:I  age ratio of about 6 (McCray
1983). Thus, the age of the adiabatic phase is less than about 1000
years.\\  

\begin{figure*}
\vbox to142mm{\vfil 
\psfig{figure=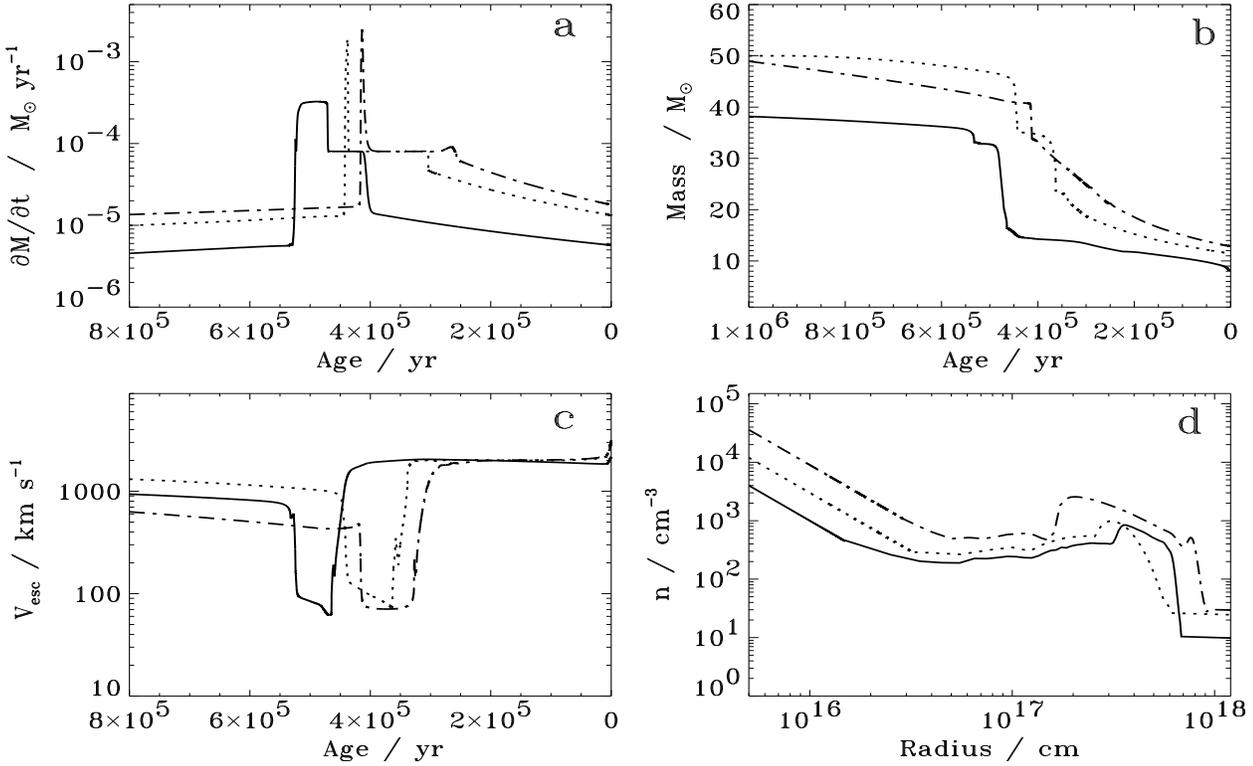,angle=0,height=115mm,width=183mm}
\caption{Stellar winds from different evolutionary WR models. The
state of the ambient medium of a WR star at the end of its life (d) is
determined primarily by the time evolution (until core collapse) of
the mass-loss 
rate (a) and the escape velocity of the wind (c). Panel (b) shows the
mass of the WR star as a function of time. Stellar models were
computed for initial masses of 
40 ({\it solid line}), 50 ({\it dotted line}) and 60 ({\it dash-dotted line})
${\rm M}_\odot$. The external medium particle
density of the wind ejected by the progenitor is best modelled (at $r
\le r_{\rm dec}$, see equation 1) by a power-law   $n=Ar^{-2}$. A
spherical shell with high column density arises when the ejected
material accumulates at the radius at which the free expansion phase
of the ejected wind terminates. We have set the density of the
ambient material before the WR phase to  $1\,{\rm cm}^{-3}$.} 
\vfil}
\label{fig1}
\end{figure*}

The mass of the
swept-up material is much larger than that in the
hot wind and,  because it is cool, it lies in a compressed region. 
Phase III persists for as long as the star is able to sustain a
powerful wind. The dominant energy loss of region II is work against
the compressed region III (see Castor et al. 1975). The
compressed region III expands because its gas pressure is higher than
that of the surrounding ISM. Therefore, the expansion is described by
the momentum equation, 
\begin{equation}
{d \over dt}[{M_{\rm S}(t)v(t)}]=4\pi r^2(t) P_{\rm i},
\end{equation}
where $P_{\rm i}$ is the internal pressure of the compressed
region, assuming that  most of the swept-up interstellar mass remains
in the thin shell. $M_{\rm S}(t)$ is the mass of the shell of swept-up
material, given by $M_{\rm S}(t)=(4/3)\pi r^3(t) \rho_0$. $P_{\rm i}$ is
determined by the gas pressure of the high temperature gas in region
II. The wind material
that enters the backward facing shock is hot, but the material that
enters the forward facing shock is cool. The cooled swept-up material is
driven outward by the high gas pressure of the hot bubble. The stellar
wind adds energy to region II at a rate $L_{\rm w}(t)={1 \over 2}\dot{M}(t)
v_{\infty}^2(t)$. The internal energy in the bubble is given by the
product of the energy per mass of the material, $(3/2)nkT / \rho_{\rm
i}=(3/2)P_{\rm i}/ \rho_{\rm i}$, and the total mass of the bubble,
$(4/3) \pi r^3 \rho_{\rm i}$. Since the total internal energy of
the bubble comes from the energy of the wind, we find 
$\dot{P}_{\rm i}=L_{\rm w}(t) / [2 \pi r^3(t)]$ (Castor et al. 1975;
McCray 1983). The expansion  of the 
bubble during the adiabatic  phase can be found numerically by
using this expression in the momentum equation\footnote{If the wind power 
$L_{\rm w}$ is roughly constant for a period of time, $t$, one can write
$P_{\rm i}={L_{\rm w}\;t / (2 \pi r^3)}$. The resulting solution of
equation (3) gives $r(t) \propto t^{3/5}$. This shows that the shell
expands more slowly than would  a freely expanding wind.}. The
bubble could continue to expand until stalled by the pressure of the
ISM. However, before that happens, the star will have produced a
supernova type explosion which re-pressurizes the bubble. 

\subsubsection{Wind-wind interaction}
During the different mass-loss stages
within a WR lifetime, the wind power of the star increases or
decreases (as shown in Fig. 1a) with time. When the escape velocity of
the wind increases, the fast wind collides with the early ejecta
before  reaching $r_{\rm dec}$. The resultant shell will be pushed
outward by the central star wind and retarded by the early ejected
wind, quickly reaching a constant velocity, but increasing in
mass. The basic theory for the wind-wind interaction was developed by
Kwok, Purton $\&$ Fitzgerald (1978). The resulting expansion law for
the shell follows from the momentum balance between the two
components (denoted as $i$ and $j$). If $M_{ij}$ is the mass of the resulting
shell when it is at a radial distance $r_{ij}(t)$, then
\begin{equation}
M_{ij}(t)=\int_{r_{j} + v_{j}t}^{r_{ij}(t)} { \dot{M}_j \over v_j}dr
+ \int_{r_{ij}(t)}^{r_{i} + v_{i}t}{ \dot{M}_i \over v_i}dr,  
\end{equation}
where $v_j < v_i$. If $v_{ij}(t)$ is the velocity of the shell,
then, assuming a completely inelastic collision, the equation of
motion may be written 
\begin{equation}
M_{ij}(t){d\;v_{ij} \over dt} =  {\dot{M}_i \over v_i}[v_i - v_{ij}]^2 -
{\dot{M}_j \over v_j}[v_j - v_{ij}]^2.  
\end{equation}
Numerical integration of equation (5), with a substitution for
$M_{ij}(t)$ from equation (4), gives the resulting expansion law for
the shell. The thickness of
the shell $\Delta r_{ij}$ may be found by requiring its internal
pressure to balance the pressure from the wind (see Kwok et
al. 1978).

\subsection{The effects of mass-loss during the evolution of WR stars}

An increase of the mass-loss rate has profound consequences for the
modelling of the WR population. At a given metallicity, the
minimum initial mass for the formation of a WR star (lower for higher
$\dot{M}$), the duration of the WR stage (greater for higher
$\dot{M}$), the times spent in the different WR subtypes and the
surface composition during these phases are all
very sensitive to the mass-loss rates. A detailed discussion is given
in Maeder (1991). Here we emphasise some interesting contributions to the
structure of the ambient  medium that arise when the ejected winds
interact  with the ISM during the WR lifetime (about $10^5 {\rm yr}$). 

\begin{figure}
\vbox to130mm{\vfil 
\psfig{figure=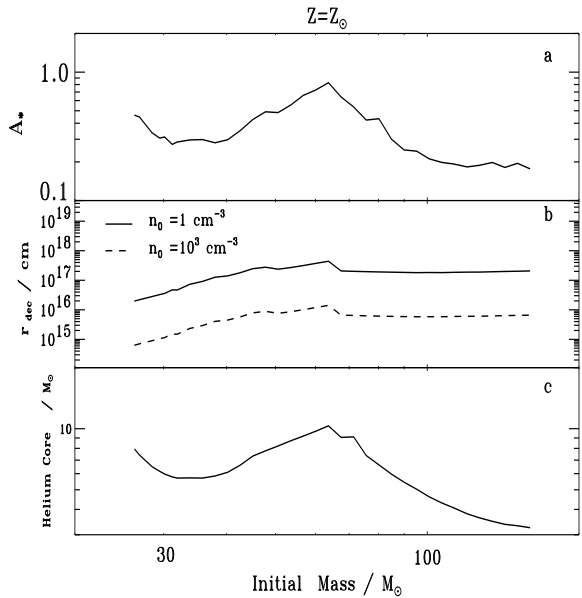,angle=0,height=80mm,width=80mm}
\caption{The effects of different initial masses on the
structure of the ambient medium. The density profile can be fit by a
power law $n(r)=Ar^{-2}$ for $r < r_{\rm dec}$. The value of the constant
$A$ as a function of $M_{0}$ is shown in panel (a) for the
stellar models  computed  in $\S$ 2.1. $A$ was
scaled to $A$=3.0$\times$10$^{35}\,A_*~{\rm cm}^{-1}$. The radial
location of the highest density shell as a function of $M_{0}$ is shown
in panel (b) for different values of $n_0$. The free expansion of the
ejected wind is terminated when the swept-up mass becomes comparable
to the mass of the wind. The ejected mass then accumulates at such a
radius ($r_{\rm dec}$). This overdense region lies
closer to the progenitor for WR stars originating from low initial
mass stars. All the evolved WR stars have lost their hydrogen envelope
and are left with a bare helium core (see panel c).}
\vfil}
\label{fig2}
\end{figure}

Using the physical
ingredients mentioned in $\S$ 2.1, we have computed the evolution
of stars with initial main-sequence masses, $M_{0}$, between
10 and 150 ${\rm M}_\odot$ and of solar metallicity, ${Z}_\odot$=0.02. 
Fig. 1a shows the time evolution (until core collapse) of the
mass-loss rate for  $M_{0}$ = 40, 50, and 60 ${\rm M}_\odot$.
The  progenitor star loses a large fraction of its initial mass quite
rapidly on entering the WNL phase (the highest constant mass-loss rate
interval in Fig. 1a). The main reason is the non-negligible contribution
of the H-burning shell to the total luminosity. Depending somewhat on
the steepness of the hydrogen profile above the H-burning shell, the
WNL stage is rather short in low initial
mass stars; its duration increases towards higher masses and
it may last for the whole post-main-sequence phase for the highest
masses \cite{langer87}. Due to the existence 
of a H-burning shell and also because 
hydrogen has a high radiative opacity and a low mean molecular
weight, the radii of WNL stars are considerably larger (and their
surface temperatures correspondingly smaller) than WNE and WC stars. 
The escape velocity of the
wind therefore decreases during this phase, as shown in Fig. 1c. 
During the early
phase of WR evolution, the progenitor stars lose a
large fraction of their initial mass quite rapidly, as shown in
Fig. 1b. We have studied the 
interaction of the strong stellar winds with the surrounding
interstellar gas and have found that a dense circumstellar shell
develops at the radius at which the free expansion phase
of the ejected wind terminates, as illustrated in Fig. 1d. For a given 
metallicity,  this overdense region is more extensive
for higher initial masses.\\

\begin{figure*}
\vbox to142mm{\vfil 
\psfig{figure=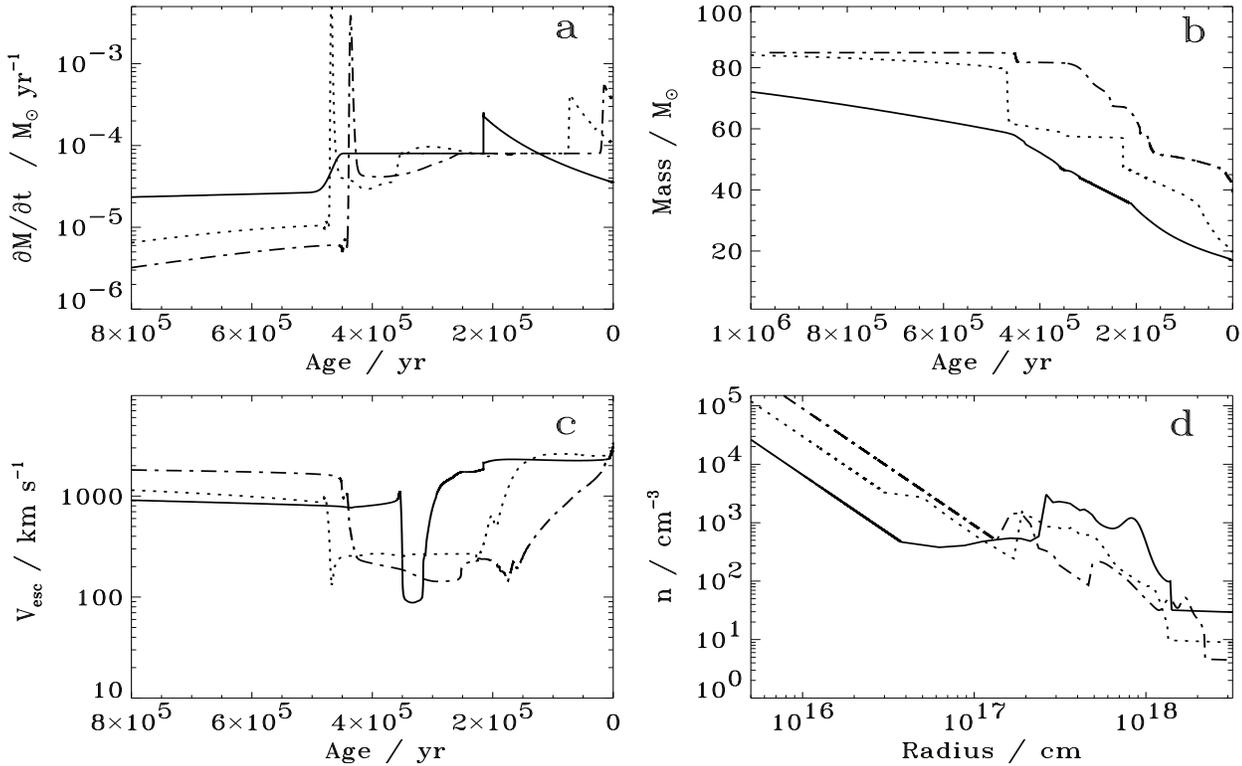,angle=0,height=115mm,width=183mm}
\caption{The effect of metallicity on stellar winds. An 85 ${\rm
M}_\odot$ main sequence star was evolved with 
three different metallicities: $Z$ = 0.5 $Z_\odot$ ({\it solid line}),
$Z$ = 0.05 $Z_\odot$ ({\it dotted line}), $Z$ = 0.015 $Z_\odot$ ({\it dash-dotted
line}). Low metallicity decreases the mass-loss rate (a), increases
the mass of the helium core (b) and keeps the
radius of the star small, reducing the escape velocity of the
stellar wind (c). Also, low metallicity favours 
high mass-loss rates at the end of the WR phase, increasing the
density of the ambient medium (d) at $r \le r_{\rm dec}$. A spherical
shell with high column density arises when the ejected wind interacts
with the ISM ($n_0 = 1\,{\rm cm}^{-3}$) and decelerates.  
In the case of the 85 ${\rm
M}_\odot$ main sequence star with $Z$ = 0.015 $Z_\odot$ a thin shell
is present at  $r \le r_{\rm dec}$ due to the substantial change in
mass-loss prior to core collapse. }
\vfil}
\label{fig3}
\end{figure*}

In most early discussions
\cite{mes98,che99,echo,pk00}, the progenitor star was expected to be
surrounded by a substantial 
\begin{equation}
n(r)=A r^{-s} 
\end{equation}
medium at the end of its life (with  $s$ = 2 for a wind ejected at a
constant speed). For the 
stellar models computed here, we found no significant deviations from
an $r^{-2}$ density gradient as a result of the
small mass-loss variations from the progenitor star prior to core
collapse. We calculated power-law indices, $s$, with values ranging from 1.9
to 2.1. By running stellar models with a variety of 
initial masses, we found that the
constant $A$ in the density profile is not a strong
function of  $M_{0}$, as shown in Fig. 2a. The constant $A$ in
equation (6) was
scaled to $A$ = 3.0 $\times$ 10 $^{35}\,A_*~{\rm cm}^{-1}$ as in Chevalier \&
Li (2000). The highest deviation from the extrapolated density
profile  is mainly provoked by the
interaction between the winds and the interstellar medium. As time
progresses the amount of material sweptup increases and the momentum
from the wind cannot drive the shell to such high speeds. The ejected
mass then accumulates at a corresponding radius, creating an overdense
region. We 
found the radial location of the overdense region or bump  to be smaller 
for WR stars that originate from low initial mass stars (see Fig.
2b). The radius at which  the free expansion ends is a
function of the product $\dot{M}^{1/2}v_\infty^{-1/2}n_{0}^{-1/2}$, so that
one may fear that large variations of $n_0$ can be masked or mimicked by much
smaller variations of $M_{0}$ (or $\dot{M}^{1/2}v_\infty^{-1/2}$; see
Fig. 2b). It should be remarked that the plausible range of $n_0$
(0.1 - $10^5$ ${ \rm cm}^{-3}$) is large compared to the 
plausible range of $M_{0}$ ($M_{0} \ge$ 35 ${\rm
M}_{\odot}$ if GRBs originate from rapidly rotating massive stars with
large helium cores, Fig. 2c).

\subsection{The role of metallicity}

Metallicity $Z$ influences the stellar evolution of massive stars
mainly through bound-free and line opacities. The
main source of 
opacity in the interior is electron scattering, which is independent of
metallicity. Therefore, metallicity has little direct structural
effect in massive stars. Bound-free and line opacities, on the other hand, are
important in the outer layers of massive stars. In this way,
metallicity influences the mass-loss
rates by stellar winds \cite{rev86}.
Low metallicity keeps the radius of the star smaller
and also reduces the 
mass-loss. Both effects inhibit the loss of angular momentum from the
evolving star (MW99). For a given
mass-loss rate, the lower the metallicity, the higher the stellar mass for
WR formation. Furthermore, low metallicity raises the
threshold for the removal of the hydrogen envelope by stellar winds,
thus increasing the mass of the heaviest helium core and favouring black
hole formation (MW99).\\

We have computed the evolution
of stellar models for an initial main sequence mass of 85 ${\rm M}_\odot$
at three different metallicities: $Z$ = 0.5 $Z_\odot$,
$Z$ = 0.05 $Z_\odot$ and $Z$ = 0.015 $Z_\odot$ (see Fig. 3). As
discussed above,  low
metallicity reduces the mass-loss for most of the WR lifetime. However,
low metallicity favours high mass-loss rates at the 
end of the WR evolution (see Fig. 3a). This increases the density of
the ambient medium at small radii prior to core collapse (see
Fig. 3d). Nevertheless, 
the overall amount of ejected
material that accumulates at $r_{\rm dec}$ increases with metallicity, as
shown in Fig. 3d.

By running stellar models with a variety of main-sequence initial
masses and  metallicities, we found that the normalization constant of
the density profile, $A_*$, is not a strong function
of $Z$, as shown in Fig. 4a. The highest deviation from the
extrapolated profile occurs at progressively smaller radii for WR
stars originating in low metallicity environments (see Fig. 4b). In
such environments, the formation of massive helium stars, desired
for GRB formation,  is favoured (see Fig. 4c).\\

The numbers of WR stars in galaxies of different
metallicities are an important test of the stellar models at various $Z$ and
of the values of the final stellar masses. The
theoretical predictions of the ratios of WR/O-stars, WC/WR and
WC/WN are in very good agreement with observations of the Local
Group \cite{m91}. This supports the adopted
dependence of the mass-loss rates on metallicity \cite{meynet94}. At
high $Z$ (in inner galactic locations), gas opacities are larger in the
outer stellar layers and so more momentum is transferred by
radiation pressure, mass-loss is more intense
and thus GRB formation is disfavoured. The strong dependence of massive star
evolution on  metallicity suggests that GRB progenitors in inner galactic
locations may be intrinsically
different from those in outer ones. Furthermore, this dependence on $Z$
implies that GRB characteristics could be strongly affected by
redshift (MW99).\\

\subsection{The effects of WR rotation}

If long GRBs  are SN-like events, the   
explosion is likely to be the outcome of black hole formation through the
collapse of a rapidly rotating massive star (WM99; Heger, Langer
\& Woosley 2000). Rotation may affect
WR evolution in several ways (Maeder \& Meynet 2000, and references
therein). First, in the case of fast rotation, the star may enter the
WR phase while still burning hydrogen in its core and thus spend more time
there. Second, rotation (through its effects on both the
mass-loss rates and mixing) may favour the formation of WR
stars with lower initial mass. Typically the
minimum mass for WR formation is 35-40 ${\rm M}_\odot$  for non-rotating
models. It decreases to
about 25 ${\rm M}_\odot$ for initial $V_{\rm rot}$ = 300 km s$^{-1}$
\cite{mm87}. Finally, higher rotation-velocities lead to longer WR
lifetimes and higher mass-loss rates. 
Thus, rotation could alleviate the need to enhance the
mass-loss rates in order to reproduce the
observed WR/O-star number ratio as originally proposed by Meynet  et
al. (1994).\\

\begin{figure}
\vbox to130mm{\vfil 
\psfig{figure=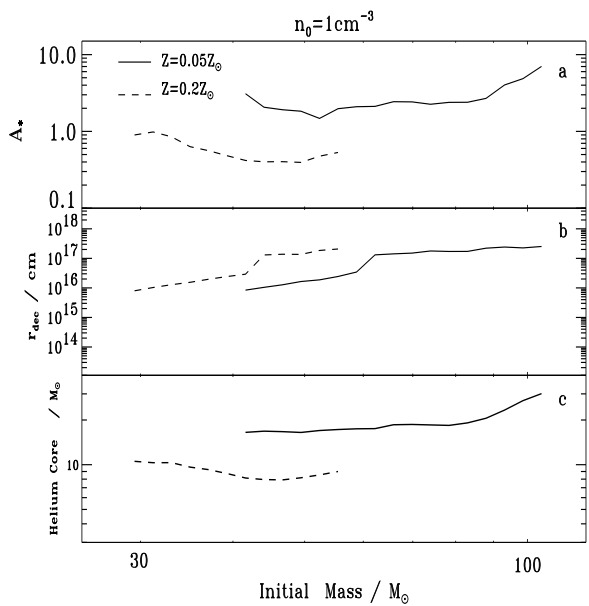,angle=0,height=80mm,width=80mm}
\caption{Stellar winds from WR models with different metallicity. All
the evolved WR stars have lost their hydrogen envelope. 
The value of the normalization constant
$A_*$ as a function of $M_{0}$ is shown in panel (a) for
different values of $Z$. The highest density shell  lies
closer to the progenitor for WR stars originating from lower $Z$
(see panel b). Low metallicity raises the threshold for removal of the hydrogen
envelope by stellar winds, increasing the mass of the helium core (panel c).}
\vfil}
\label{fig4}
\end{figure}

Direct attempts to measure the rotation velocity of WR stars have
been performed only for a few cases: Massey (1980) obtained
$v {\rm sin}i \sim {\rm 500\;}{\rm  km\; s}^{-1}$ for WR138
(see also  Koenigsberger 1990); Massey \& Conti
(1981) measured  $v {\rm sin}i \sim {\rm 150 - 200\;km\; s}^{-1}$ for
WR3. Also, indirect evidence has been found for the existence of
some axisymmetric features around WR stars. For example, 
Harries et al. (1998) suggest that about 15 per cent of WR stars have
anisotropic winds caused by equatorial density enhancements produced by rapid
rotation. High rotation velocities for iron cores 
favour the collapsar model for GRBs (see MW99). 
The contribution of the
WR mass-loss to the surrounding medium along the GRB beam,
which is likely to be close to the rotation axis of the star, may be smaller
than calculated here for the spherical wind case.

\section{Wind Interaction Models for GRB Afterglows}

One can understand the dynamics of the afterglows of GRBs in a fairly
simple manner, independent of any uncertainties about the progenitor
systems, from the relativistic generalization of the method used to
describe supernova remnants. The basic model for GRB afterglow
hydrodynamics is a relativistic blast wave expanding into the
surrounding medium \cite{mes97}. The interaction of the outer shell with
the external medium is described by the adiabatic Blandford-McKee
(1976, hereafter BM) self-similar solution. The scaling laws that are
appropriate for the burst interaction with a medium with particle density
$n \propto r^{-s}$ have been described by M\'esz\'aros et al.
(1998), Chevalier \& Li (1999,2000), and Panaitescu \& Kumar (2000). We
aim to examine the specific predictions for the interaction of
the relativistic blast wave with the density bumps that arise when 
ejected stellar material decelerates against the external medium 
(see $\S$ 2). This collision  would be an important
contribution to the observed afterglow if it were not to take place too
late in the fireball evolution. We outline the relevant physical parameters
that determine when this collision occurs in $\S$ 3.1. 
We discuss briefly the situation when the
burst is collimated in $\S$ 3.2. We analyse the shock conditions when
the shell collides with an enhanced density medium in $\S$ 3.3, and
estimate the synchrotron emission that results and its observational
consequences in $\S$ 3.4.  
  
\subsection{Hydrodynamics of a relativistic shell}

For an adiabatic ultra-relativistic blast wave, the (isotropic
equivalent) total energy is 
\begin{equation}
E= {8\pi A \Gamma^2 r^{3-s}c^2 \over 17 - 4s},
\end{equation}
where $\Gamma$ is the
bulk Lorentz factor of the shock front and $r$ is the observed radius
near the line of sight (BM). We assume the burst to be collimated
with an initial half-angle $\theta$ larger than
20$^{\rm o}$, and that lateral expansion is negligible during the
relativistic phase. A distant observer receives a photon emitted along the
line of sight towards the fireball centre at a time
$t=r/4(4-s)\Gamma^2c$  \cite{che00}, and so
\begin{equation}
r = \left[{(4-s)(17-4s)Et \over 2\pi Ac }\right]^{1\over(4-s)}.
\end{equation}
Before the collision with the high density shell, the shock front is
expected to propagate through an $n(r)=Ar^{-s}$ wind. In
a spherically symmetric wind ejected at a constant speed the density
drops as $s$=2. Using the stellar
models in $\S$ 2, we found $ 0.1< A_* <10$ for WR stars
surrounded by an $s \approx 2$ medium at the end of their
life ($A_*$ = $A$ / 3.0$\times$ 10 $^{35}\,{\rm cm}^{-1}$). For these
stellar parameters, a
re-brightening of the afterglow as a consequence of the collision
of the shock front with the high density spherical shell is observed
at a time
\begin{equation}
t_{\rm day}\approx  0.25 \left({ E \over 10^{52}\,{\rm ergs}}
\right)^{-1}\left({r_{\rm shell}\over 10^{17}\,{\rm cm}}\right)^2 A_*
\end{equation}
after the burst. Here $t_{\rm day}$ is the
observer time measured in days. If the collision takes place at a
sufficiently late phase of the afterglow the system is adiabatic.
This is probably 
valid from about half an hour after the burst (Granot,
Piran \& Sari 1998). The outer shell catches up with the thin  high
density shell, which is at rest at a radius $r_{\rm shell}$. The
shell collision produces two new shock waves: a forward shock that
moves into the thin shell and a reverse shock that
propagates into the relativistic ejecta. The calculation of shocks in
the collision of two cold shells resembles the calculation of
energy emitted from internal shocks \cite{kpi00}.

\subsection{Collimation of $\gamma$-ray bursts}

The energy flux from a GRB may be collimated, as witnessed in the jets
from active galactic 
nuclei and some binary black hole sources. Some models explicitly
include this collimation (MW99). As long as the jet is highly
relativistic, the observed features are reproduced by spherical
models, but as the shocked jet slows down there are clear consequences
for the observed afterglow \cite{rhoads97}: the edge of the jet becomes
visible when $\Gamma=(2 \theta)^{-1}$ and the jet is able to
expand laterally, leading to an exponential slowing of the
forward shock front. In the case of a collimated  GRB going off in a
medium with density decreasing as $r^{-2}$, jet effects
are expected to become important at a time $t_{\rm jet} \approx 2(\theta/0.1)^4
E_{52}A_*^{-1}$ days \cite{che00}. Kumar \&
Panaitescu (2000), have recently pointed out, however, that a fireball
propagating into an $s=2$ medium should show little evidence for light-curve
steepening due to edge and sideways expansion. This could explain the
lack of breaks in the afterglow lightcurves of GRB 980326 and GRB
980519. In a collimated outflow the
sharpest break in the lightcurve occurs in a uniform low density
medium and is associated with the edge of the jet approaching the
relativistic beaming cone. In wind scenarios, where no
clear break in the lightcurve is predicted, jets can perhaps be
detected by time dependent measurements of polarization \cite{kp00}.\\

For a collimated outflow expanding in an  $r^{-s}$ medium, the observed
radius near the line of sight is proportional to
$(Et/\theta^2Ac)^{1/(4-s)}$. Thus, for a fixed value of $E$, the impact of the
relativistic shell with a density bump would  take place earlier
in the evolution of the fireball if the burst is collimated, causing
a re-brightening of the afterglow to appear at earlier times in the
observer's frame.  
 
\subsection{Shock conditions} 

Consider a relativistic shell with a Lorentz
factor $\eta$ moving into a cold
external medium (EM). The interaction is described by two shocks: a
forward shock  propagating  into the EM and a reverse shock
propagating into the shell. There are four regions 
separated by the two shocks: the EM (1), the shocked EM (2), the
shocked shell material (3) and the unshocked shell material (4). The
EM is at rest relative to the observer. From the jump conditions for
the shocks and the equalities of pressure and 
velocity along the contact discontinuity we can estimate the Lorentz
factor $\gamma$, the number density $n$ and the energy density $e$
in the different shocked regions as functions of the three quantities 
 $n_1$, $n_4$ and $\eta$ (BM).
The unshocked EM has $\gamma_1$ = 1, while the
unshocked shell material moves at the original coasting velocity of
the particles, $\gamma_4=\eta$.  
The afterglow emission is determined by the deceleration time
of the shell. The emitting region moves toward the observer with a Lorentz
factor $\Gamma=\gamma_2=\gamma_3 $, assuming that the shocked material
(either region 2 or 3) emits radiation on a timescale shorter than
the hydrodynamical time. \\

There are two limits in which there is a simple analytical
solution \cite{sari95}. If the shell density is high, $n_4 \gg
\eta^2 n_1$, then the reverse shock is Newtonian (or mildly relativistic). In
this case the energy conversion takes place in the forward shock, and
the collision is too weak to slow down the shell efficiently,
$\Gamma \sim \eta$. If the density of the
expanding shell is low, $n_4 \ll \eta^2 n_1$, then the reverse shock
is relativistic, the shell material decelerates considerably and 
\begin{equation}
\Gamma \sim (n_4 /n_1)^{1/4}(\eta /2)^{1/2}.
\end{equation}
The energy density in the shocked regions satisfies $e=e_2=e_3=4\Gamma^2
n_1m_{\rm p}c^2$. Similar relations hold for the reverse shock \cite{sprs}. 
In any realistic
situation the EM is probably inhomogeneous, 
as in the stellar models in $\S$ 2.\\

Consider a density
jump by a factor $\alpha$ over a distance $l$. The forward shock
propagates into the EM with a density $n_1$ as before, and when it
reaches the position where the EM density is  $\alpha n_1$ a new shock
wave is reflected. This shock is reflected again off the shell. 
Sari \& Piran (1995) showed that the reflection
time is $\approx l/ (4c\alpha^{1/2})$. After these reflections the
Lorentz factor and hydrodynamical properties of the system are as if
the EM were 
homogeneous with a density $\alpha n_1$. The corresponding
observed timescale due to this inhomogeneity of the EM
is $t \approx l/(\alpha^{1/2}\Gamma)$ (for $l \ga r$).

\subsection{Synchrotron emission}

The synchrotron spectrum
from relativistic electrons that are continuously accelerated into a
power-law energy distribution comprises four power-law
segments, separated by three critical frequencies, the self-absorption 
frequency ($\nu_{\rm sa}$), the cooling frequency
($\nu_{\rm c}$) and the characteristic synchrotron frequency ($\nu_{\rm m}$)
\cite{spn98,mes98}. The spectrum and light-curve of an afterglow
are determined by the time evolution of these frequencies, which in
turn depend on the hydrodynamical evolution of the fireball.

\begin{figure}
\vbox to145mm{\vfil 
\psfig{figure=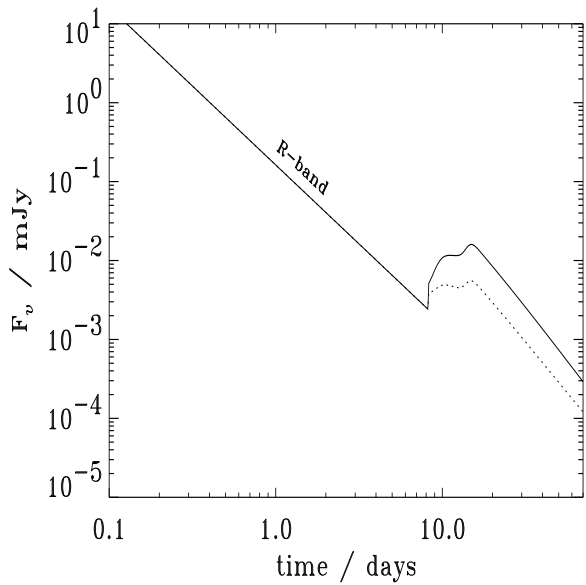,angle=0,height=90mm,width=90mm}
\caption{The effect of the impact of a relativistic shell with
the density discontinuity on the optical afterglow. The shock front
expands within the $s=2$ stellar wind until it reaches the high
density shell at a distance $r_{\rm shell} \approx 4 \times 10^{17}$cm. The
radial profile of the high-density shell is modelled as the inner
bump shown in Fig. 1d for the 40 ${\rm M}_\odot$ case. 
Collisional models were computed for
a high density enhancement $\alpha$=100 ({\it solid line})
and a low density enhancement $\alpha$=10 ({\it dotted line}). 
At the time of the collision the
relativistic shell Lorentz factor is $\Gamma \sim 3$ for
$E_{52}$=0.5, $A_*$=1, $\varepsilon_e$=0.1, $\varepsilon_B$=1.0,
and $p$=3. The afterglow emission is calculated in the adiabatic regime. 
The collision model takes into account the
fireball geometrical curvature when calculating the photon arrival time and
relativistic boosting. Given the  radius
of collision $ r_{\rm shell} \gg l$, the observed
variability timescale is $\Delta t_{\rm obs} \approx r_{\rm shell} /
(c \Gamma^2) 
>$ 20 days. Note that $ \Gamma \ll $ 3 after the collision. } 
\vfil}
\label{fig5}
\end{figure}

The break frequencies can be calculated if the energy distribution of the
injected electrons and the strength of the magnetic field are
both known. The distribution of the injected electrons is assumed to be
a power law of index $-p$, above a minimum Lorentz factor
$\gamma_i \approx \varepsilon_{\rm e}(m_{\rm p}/m_{\rm e})\Gamma$. The energy
carried by the electrons is a fraction ${(p-1)/(p-2)}$ of the total internal
energy  $\varepsilon_{e}$. The turbulently 
generated magnetic field, assumed to be amplified up to a fraction
$\varepsilon_{B}$ by the processes in the shocked region and not 
determined directly by the field in the WR star, is $B \approx \sqrt{32\pi}c
\varepsilon_{B}^{1/2} m_{\rm p}^{1/2} n_1^{1/2} \Gamma$ \cite{spn96}. 
For an adiabatic blast wave, the
corresponding observer peak frequency is
\begin{equation}
\nu_m \propto  {B \over m_{e}c} \gamma_i^2 \Gamma.
\end{equation}    
The ratio of $\nu_{m,\alpha}$ the observed peak frequency from the
relativistic shell after the forward shock has traversed a density jump of a
factor $\alpha$ over a distance $l$, and the
frequency in the absence of such a jump $\nu_{m}$,  is 
\begin{equation}
{ \nu_{m,\alpha} \over \nu_m} \approx { \Gamma_\alpha^4 (\alpha
n_1)^{1/2} \over \Gamma^4 n_1^{1/2}} \approx \alpha^{-1/2}.     
\end{equation} 
Therefore, we expect this emission to be a very
significant contribution to the short-wavelength flux.
The spectrum at low frequencies ($\nu \ll \nu_m$) scales
as $\nu^{1/3}$; at high frequencies ($\nu \gg \nu_m$) it scales as
$\nu^{-(p-1)/2}$.  At late times, when the collision has run its
course, the shells have merged back to a BM solution and the observed
flux is proportional to $E_\alpha^{(p+1)/[2(4-s)]}$ for $n(r)
\propto r^{-s}$. In the absence of a density
enhancement, the flux would have been proportional to
$E^{(p+1)/[2(4-s)]}$ \cite{kpi00}. 
Therefore, the increase in the observed emission from the forward
shock due to the density jump is approximately
$(E_\alpha/E)^{(p+1)/[2(4-s)]} \approx \alpha^{(p+1)/[4(4-s)]} $,
neglecting the enhanced 
energy losses from the shocks that may arise during the collision with the
high-density thin shell. We should expect, however, a smooth
transition from one solution to the other at most wavelengths.  This
is not the case at 
frequencies for which there is significant emission from the reverse
shock, from which the total  emission is smaller,
and at a lower frequency by typically a factor of
$\Gamma$.  At these frequencies the impact of this additional emission
is likely to be significant.\\

We conclude that the impact of the relativistic shell with
the density discontinuity (see Figs. 1d and 3d) produces a  
bump in the lightcurve that can be very
dramatic at low frequencies. To  illustrate this, we consider the scaling
laws described by Chevalier \& Li (1999)  for an adiabatic blast wave
in an s = 2 medium. Before the collision, the synchrotron
emission frequency of the lowest energy electrons is $\nu_m=5 \times
10^{12} \varepsilon_{e,-1}^{2}
\varepsilon_{B,-1}^{1/2} E_{52}^{1/2}t_{\rm day}^{-3/2}$ Hz and the flux
at this frequency is $F_{\nu_m}=20 A_* \varepsilon_{B,-1}^{1/2}
E_{52}^{1/2}t_{\rm day}^{-1/2}$ mJy (These expressions assume  $z=1$ in
a flat Universe with $H_0=65\,{\rm km\,s^{-1}\,Mpc^{-1}}$). For
electrons with a power-law distribution $p$, the flux above $\nu_m$ is
$F_\nu = F_{\nu_m}(\nu/\nu_m)^{-(p-1)/2} \propto t^{-(3p-1)/4}$. Over
the typical time of observation of a GRB afterglow, the shock front
expands within the stellar wind until it reaches the high-density bump
at a distance of about $4 \times 10^{17}$ cm.  The
impact of the relativistic shell with the density enhancement
produces a clear feature in the afterglow lightcurve. Fig. 5 shows
the effect for two different values of the bump
density. 
The lightcurve of the enhanced emission arising from the
interaction of the forward shock with the density bump is calculated
with the contribution of the curvature to the temporal structure
described in Fenimore, Madras \& Nayakshin (1996). Given a typical radius
of collision $r_{\rm shell} \gg l_{\rm shell}$, the observed
variability timescale is $\Delta t_{\rm obs} \approx  r_{\rm
shell}/(\Gamma^2c)$.     
The maximum
enhancement of the observed emission with respect to the extrapolated
afterglow lightcurve is approximately $
\alpha^{1/2}$ and the characteristic synchrotron
frequency is lower than that in the absence of
the collision.

\section{Compton echoes}

If massive stars are the progenitors of GRBs, then the hard photon pulse
will propagate in a pre-burst stellar wind. The GRB is assumed
to have a broken power-law spectrum, $\epsilon E_\epsilon \propto
\epsilon$ ($\epsilon\le 250\,$ keV) and $\epsilon^{-0.25}$
($\epsilon>250\,$ keV). Circumstellar
material Compton scatters the prompt radiation and gives rise to a
reflection echo \cite{echo}. If $E = \int
E_{\epsilon\Omega}d\epsilon d\Omega$ is the total energy emitted by the 
burst, where $E_{\epsilon\Omega}$ is the energy emitted per unit energy 
$\epsilon$ and unit solid angle $\Omega$ along the $\theta$ direction, 
then the equivalent isotropic luminosity of the Compton echo
inferred by a distant observer is
\begin{equation}
L_{\epsilon'}=4\pi\int n_e(r) E_{\epsilon\Omega} {d\sigma\over 
d\Omega} {dr\over dt}d\Omega,  
\end{equation}
where $r$ is the distance from the site of the burst, $d\sigma /
d\Omega$ is the differential Klein-Nishina cross section for
unpolarised incident radiation, $\sigma_T$ is the Thomson cross
section, $\epsilon'$ is the energy of the scattered photon and  $n_e$
the local electron density. The equal-arrival time scattering material
lies on the paraboloid determined by 
\begin{equation}
r={ct\over 1-\cos \theta}\;\;,
\end{equation} 
where $\theta$ is the angle between the line of sight and the direction of the 
reflecting gas as seen by the burst.\\ 

Madau et al. (2000)
derived analytical expressions for the echo luminosity lightcurve in
the Thomson regime for a range of possible emission geometries and ambient gas
distributions. In particular, they discussed the effect of an
infinitesimal high-density spherical shell
in the surrounding medium. For isotropic emission,
the echo luminosity decreases as the equal-arrival time paraboloid
sweeps up the shell, reaching a minimum when $t=r_{\rm shell}/c$, and 
then increases until it reaches the back of the shell. 
For a collimated burst, the
echo lightcurve is the sum of 
two delta functions,  as the result of the interaction of the
approaching and receding jet with the spherical shell (see equation 10 in  
Madau et al. 2000). 
The two spikes of emission are seen separated
by an interval 2$r_{\rm shell} \cos \theta /c$, where $\theta$ is the
angle between the line of sight and the approaching beam (i.e., the
main burst emission is not seen by the observer).\\

\begin{figure}
\vbox to115mm{\vfil 
\psfig{figure=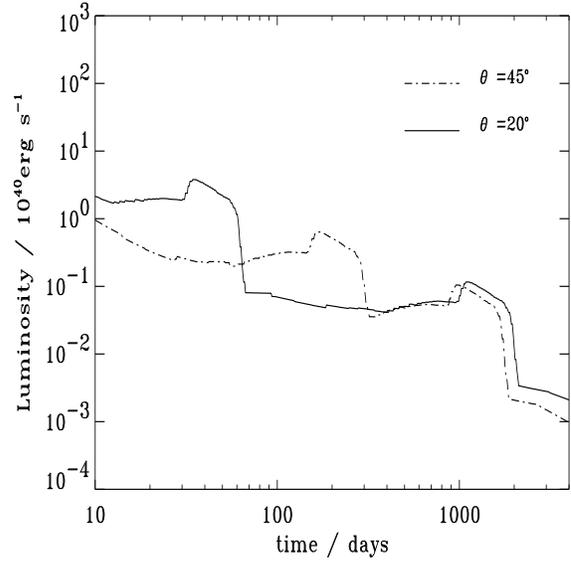,angle=0,height=90mm,width=90mm}
\caption{The Compton echo of a GRB. The primary burst is assumed to be 
a two-sided collimated pulse of energy $10^{52}$ ergs s$^{-1}$ and
duration 10 s, propagating at an angle 
$\theta$ to the line of sight, and invisible to the observer. The
GRB jet propagates through the dense 
environment expected at the end of the evolution of a 40 ${\rm M}_\odot$
Wolf-Rayet star with $Z_\odot$ (see Fig. 1d).} 
\vfil}
\label{fig6}
\end{figure}

Fig. 6 shows
the reflected echo of a two-sided GRB jet propagating through the dense
environment expected at the end of the evolution of a 40 ${\rm M}_\odot$
Wolf-Rayet star with $Z_\odot$ (see Fig. 1d). We calculated the luminosity of
the reflected echo by integrating equation (13) with the
full Klein-Nishina cross section. The GRB is assumed to radiate a
total of 10$^{53}$ erg, with each jet having equal strength. On the
equal-arrival time paraboloid, the receding beam is reflected by
denser gas that is closer to the source, and so  its contribution
dominates the echo at all energies where the scatter occurs in the
Thomson regime. This is clearly illustrated by the second spike of the
echo caused by the interaction of the receding
jet with the density bump in the surrounding medium. The
spectral energy distribution mirrors the prompt burst at low energies,
but is much steeper above a few hundred keV (Madau et al. 2000). 
The hard X-ray flash from
a pulse propagating into the environment expected at the end of the
evolution of a 40 ${\rm M}_\odot$ star could be detectable by {\it Swift} out
to z $\approx$ 0.1.
It is instructive to look at the contribution of the density bump to the
luminosity of the reflected echo. The time interval between the the two
spikes, 2$r_{\rm shell} \cos \theta /c$, can be used to determine the angle
between the line of sight and the approaching
beam.

\section{Recent observations and future predictions}

\subsection{Afterglow Sources}

The variety of observed afterglows, while
compatible with relativistic fireball models \cite{mes98}, poses 
challenges for interpretation. Some of the key questions
include the effect of the external medium, the possible anisotropy of
afterglow emission and the
radiative efficiency. In
this paper, we have considered the effects of the surrounding medium
due to massive WR stellar progenitors on GRBs. We found that a high-density
bump arises when the ejected wind interacts with the external medium
and decelerates, or when the progenitor star rapidly loses a large
fraction of its initial mass. The impact between the relativistic
shell and this density bump could be observed as  a
re-brightening\footnote{Note that a 
re-brightening can also be produced by shock refreshment due to 
delayed energy injection by an extremely long-lived central engine. 
In this case, however, the spectrum would be
bluer than  a typical GRB afterglow (see Panaitescu, M\'esz\'aros \&
Rees  1998 for the case of GRB
970508).} of the afterglow with a
spectrum typically redder than the synchrotron afterglow spectrum, as
seen in GRB 980326 and GRB 970228. These observations have been
explained as an underlying supernova  outshining the afterglow some
days after the burst event. However, this 
interpretation should be regarded as tentative, in consideration of 
the fact that only R band measurements and a low signal-to-noise
spectrum were obtained for GRB~980326. In the case of GRB~970228 no
spectroscopic measurement was made at the time of the suggested
re-brightening, and the lightcurve was unevenly sampled in different
filters. 
 
\subsubsection{GRB 970228 \& GRB 980326}
In order to explain the re-brightening of GRB 980326, we
would require a large density jump with 
a size comparable to the variability timescale, about $10 \Gamma
\;{\rm light days} \approx 10^{17}$ cm, and located at 
a distance ${\rm 20\Gamma^2 c\;\;days} \approx 10^{18}$ cm
\cite{blo99}. These conditions are clearly compatible with the density jumps
calculated in $\S$ 2. 
A very red spectrum $F_{\nu} \propto \nu^{-2.8}$
(compared with   $F_{\nu} \propto \nu^{-1.0}$ in the extrapolated
afterglow) can be produced if there is some contribution to
the emission from the reverse shock at lower frequencies. In the case
of GRB 970228 at $z=0.695$, the isotropic 
luminosity is $6 \times 10^{51}$ erg s$^{-1}$. The temporal bump
detected 10 days after the explosion can be explained by the impact of
a  relativistic shell expanding within an
$n(r)=A_*r^{-2}$ stellar wind with the density discontinuity located at a
distance $5 \times 10^{17}$cm (see equation 9).  These
values are in agreement with the overdense regions calculated in
$\S$ 2.\\

\subsubsection{GRB 000911}

\begin{figure}
\vbox to115mm{\vfil 
\psfig{figure=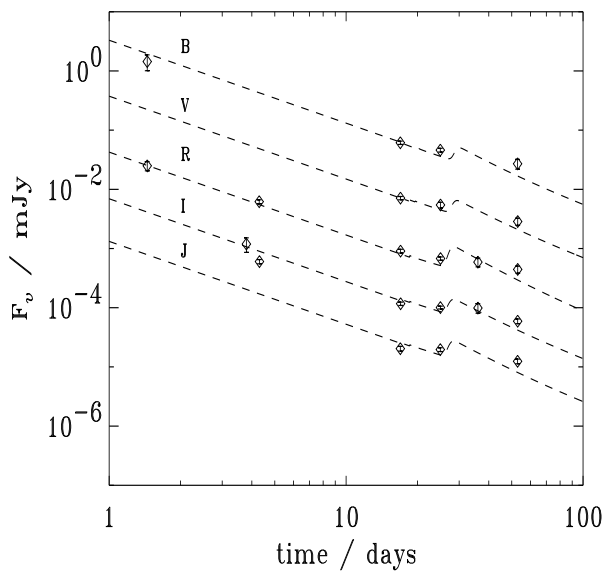,angle=0,height=90mm,width=90mm}
\caption{Modelling of the multi-filter lightcurves of the afterglow of
GRB 000911. From top to bottom the normalised photometric measurements
are plotted, scaled by factors of $10^2$, $10^1$, 1, $10^{-1}$,
$10^{-2}$ for clarity. The dashed lines show the best fit for a collisional
model obtained by computing the interaction of a relativistic blast wave
with the WR ejecta.}    
\vfil}
\label{fig7}
\end{figure}

Using detailed stellar tracks for the evolution of massive stars, as
described in $\S$ 2,  we found that  the re-brightening of
GRB 0009111 at $z=1.1$ (Lazzati et al. 2001), can be explained by the
interaction of a  relativistic blast wave ($\varepsilon_e$=0.1,
$\varepsilon_B$=0.1, and $p$=2.2) expanding into the ambient medium
expected at  the end of the life of a 40 $M_{\odot}$ (7 $M_{\odot}$ core)  
WR star evolving with $Z$=$Z_\odot$.  The shock front expands within a
$n(r)= 1.5 \times 10^{35}{\rm cm}^{-1} r^{-2}$ stellar wind until it
reaches the density enhancement (see Fig. 1d). The very red
spectrum $F_{\nu} \propto \nu^{-5}$ (compared with   $F_{\nu} \propto
\nu^{-1.4}$ in the extrapolated  afterglow) can be better reproduced
if there is a substantial contribution to the emission from the
reverse shock at lower frequencies. We have fitted this model to the
simultaneous multi-filter data (see Fig. 7), obtaining a reduced
$\chi^2$ intermediate  between the fit with a simple  external shock
synchrotron component plus a  galaxy model ($\chi^2 \sim$ 1.3)  
and the fit with the additional SN component ($\chi^2 \sim$ 1.1, see
Lazzati et al. 2001).

\subsection{Internal shock models}

In the internal shocks scenario for GRBs, the actual $\gamma$-ray 
temporal profile is the
outcome of the complex dynamics of the ejecta. A growing consensus is
that a central site releases energy in the form of a wind or multiple
shells over a period of time commensurate with the observed duration
of GRBs (Rees \& M\'esz\'aros 1994). Each subpeak in the $\gamma$-ray
lightcurve is the result of a separate explosive event at the central site. 
In this early phase, the
time-scale of the burst and its   
overall structure follow, to a large extent, the temporal behaviour of
the source.
In contrast, the subsequent afterglow
emerges from the shocked regions of the external medium where the relativistic 
flow is slowed down and the inner engine cannot be seen
directly. A concern was raised that internal shocks
without deceleration were rather inefficient (Kobayashi, Piran \& Sari
1997) converting at most 25 per cent of  
the bulk motion energy into radiation. Since the afterglows can only
account for a few percent of the radiated energy, it is unclear
where most of the energy goes.\\ 

In a dense molecular cloud  the
density of the surrounding medium, before the free expansion phase, can be as
large as  $n_{0} \approx 10^4-10^5 {\rm cm}^{-3}$ (Shull 1982;
Sanders, Scoville \& Solomon 1985)    
This is not unlikely because massive stars tend 
to cluster and  these stars do not live long enough to travel far from
their place of birth. In this environment the free expansion phase of the
ejected wind, which  is terminated when the swept-up mass becomes
comparable to the mass in the wind, ends at a
radius of $10^{15}-10^{16}$ cm (see equation 1). The ejected mass then
accumulates at such a radius, creating an overdense
region. Substantial internal energy can be converted into radiation at
this radius when the shells 
responsible for the $\gamma$-ray emission run into this region. This is 
equivalent to the model proposed by Fenimore \& Ramirez-Ruiz (2001),
with the dense region acting as the decelerating shell. No substantial
deceleration is required to convert
up to 80 per cent of the bulk-motion energy during the GRB. Indeed, a recent
analysis of 387 pulses in 28 BATSE GRBs shows that the most
intense pulses within a burst have nearly identical widths throughout the
burst, but that the weak pulses tend to become wider as the burst
progresses \cite{err00}. This effect is seen in internal
shock simulations when a high density thin shell at a radius of
$10^{15}-10^{16}$ cm is included. Furthermore, the detection of an
optical flash associated with GRB 990123 \cite{rotse99}, with a peak
magnitude of 9 in the V band 50 s after the initial burst (successfully
described by the reverse shock model by Sari \& Piran
1999), can also be explained as a result of the impact of two shells that
collide with a relative Lorentz factor of the order of 2
\cite{kpi00}. Such collisions are expected in internal-shock
models if an ejected shell catches up with a decelerating shell or when 
a shell runs into a dense region at a radius of $10^{15}-10^{16} $ cm.

\section{The $\gamma$-ray burst environment}

The task of finding useful progenitor diagnostics is simplified if the
pre-burst evolution leads to a significantly enhanced gas density in the
immediate neighbourhood of the burst. The detection of spectral
signatures associated with the GRB environment would
provide important clues about the triggering mechanism and the
progenitor. Stars interact with the surrounding interstellar
medium, both through their ionising radiation and through
mass, momentum and energy transfer in their
winds. Mass-loss leads to recycling of matter
into the interstellar medium, often with chemical enrichment. 
Mass-loss is a particularly significant
effect in the evolution of massive stars; in particular, for stars
that enter WR stages. WR stars follow all or part of the sequence WNL, WNE, WC
and WO, which corresponds to a progression in the exposure
of nuclear products: CNO equilibrium with H present; CNO
equilibrium without H; early visibility of the products of the
3$\alpha$ reaction;  and then a growing (C+O)/He ratio respectively. 
Mass-loss drastically influences
stellar yields. At low $Z$ there is a high production of
$\alpha$-nuclei, while at higher $Z$ most of the He and C produced is
ejected in stellar winds and escapes further nuclear
processing.\\ 

The environment could also be metal-enriched, although solar
abundances are probably more likely. The amount of metals in
the environment can be  larger than the stellar yields  if a supernova
explosion occurs before the GRB, perhaps producing 
significant Fe K$\alpha$ and K-edge luminosities \cite{davide}. 
Because of the very high
luminosities and hard initial $\gamma$-ray spectrum, all the Fe  
is fully ionised, and as it cools, the strong  
Fe K$\alpha$ and K-edge appear initially in absorption,
and later as recombination emission features \cite{weth00}. Guided by
the report of  Fe line detections that peak after 1 day in GRB 970508
and GRB 970828 \cite{piro99,yosh99}, we are led to consider a high
density shell. A physical requirement is
that the distance $ct\Gamma(t)^2$, reached after 1 day by the afterglow
shock, is less than the shell radius. Notice that in order to reproduce
the observed Fe K$\alpha$ equivalent widths, a shell located at few $10^{16}$ cm
is required with a mass of Fe $\sim 2.5 \times 10^{-4} M_\odot$, in a
total shell mass of 1 $M_\odot$ (Weth et al. 2000). Such enhanced
density shells can be present in the environment of a progenitor star 
if the density of the surrounding
medium, before the WR phase, is of order  $10^5 {\rm
\;cm}^{-3}$. The ejected mass accumulates at
this radius, creating a high-density region 10$^{9}$-10$^{11}$
cm$^{-3}$, for a 50 ${\rm M}_\odot$ main-sequence star. Without
substantial deceleration of the wind 
ejecta, such high density shells are difficult to form. Emission lines
are hard to detect, unless the event
occurs in an exceptionally dense or metal-rich environment (Piro et
al. 2000). A massive star environment could be distinguished by the presence
of a significant flux of  Fe K$\alpha$ and probably H
Ly-$\alpha$ emission lines, reprocessed by a moderately Thompson
thick envelope. For a more massive thick envelope,
a significant reflected component would be expected, in which Fe
absorption edges and K$\alpha$ features would be present
\cite{edge98,amati00}.

\section{Conclusions}

Recent observations strongly suggest that at least some GRBs are
related to massive supernovae. Quantitative insight
into the formation of GRBs is hindered
by the lack of detailed core-collapse calculations. The
ground-breaking  
work of MW99 suggests that GRBs are more
likely to occur in stars that have lost their hydrogen
envelope. Stars with less radiative mass-loss retain a hydrogen
envelope in which a poorly
collimated jet that loses its energy before breaking through the stellar
surface is likely to arise. Highly relativistic
jets will not escape red supergiants with radii $> 10^{13}$
cm. By contrast, a focused low-entropy jet that
has broken free of its stellar cocoon is likely to arise from a WR
progenitor. Our
detailed stellar tracks for  the evolution of a WR
star show the diverse effects that initial main-sequence mass,
metallicity, and rotation can have on the
surrounding medium.  Pre-explosion mass loss
provides a natural medium to generate GRB afterglows, and allows
specific predictions about their temporal 
structure to be made. The presence of density bumps in the
nearby ambient environment are inherent to the evolution of WR stars. The
impact between the forward shock and these high-density regions should
be observed
as a re-brightening of the afterglow with a typically redder
spectrum. Interestingly, the density ring in the wind profile lies
closer to the progenitor
for WR stars with low initial mass and low metallicity. 
This characteristic offers a direct observational test of which
stars are likely to produce a GRB. \\ 

The total
energy observed  in $\gamma$-rays from GRBs  whose redshift has been
determined is diverse. One appealing aspect of a massive star
progenitor is that the great variety of stellar parameters can probably
explain this diversity.  Given the need for a large helium core
mass in progenitors, the burst formation may be favoured not only by rapid
rotation but also by low metallicity. Larger mass helium cores might have
more energetic jets, but it is unclear whether they can be expected to
be accelerated to large Lorentz factors. Many massive stars may
produce supernovae by forming neutron stars in 
spherically symmetric explosions, but some may fail neutrino energy
deposition,  forming a black hole in the centre of the star and
possibly a GRB. One expects various outcomes ranging from GRBs
with large energies and durations, to asymmetric, energetic supernovae
with weak GRBs. The medium  surrounding a GRB would provide
a natural test to distinguish between different stellar explosions.

\section*{Acknowledgments}

We thank P. Natarajan, A. Celotti, G. Morris, G. Koenigsberger,
A. MacFadyen and the referee for useful  comments and suggestions. We
are particularly grateful to A. Blain, R. Chevalier, D. Lazzati,
M. Rees  and R. Wijers for helpful conversations. ER-R acknowledges
support from CONACYT, SEP and the ORS foundation. LMD acknowledges
support from PPARC.

\bsp

\label{lastpage}

\end{document}